\RequirePackage{fix-cm}
\RequirePackage{fix-cm}
\documentclass[11pt,notitlepage,a4paper]{article}
\usepackage[a4paper,left=2.84cm,right=2.84cm,top=2.96cm,bottom=2.96cm,footskip=1.5cm]{geometry}
\renewcommand\footnotemark{}
\makeatletter
\renewcommand*{\@makefnmark}{}
\makeatother
\usepackage[hang,flushmargin]{footmisc}
\usepackage[hidelinks]{hyperref}
\usepackage{tocloft}

\newlength{\mylen}	
\usepackage{amsmath,amssymb,amscd,amsfonts}
\usepackage{enumitem}

\newcommand\onlinecite{\cite}
\newcommand\kB{k}

\newcommand{\dif}{\mathrm{d}}
\newcommand\rhoent{\rho_{S}}
\newcommand\eq{\begin{equation}}
\newcommand\en{\end{equation}}
\newcommand\expval[1]{\langle #1 \rangle}
\newcommand\expvalequil[1]{\expval{#1}_{\mathit{eq}}}
\newcommand\Jent{j_{S}}
\newcommand\Phient{\Phi_{S}}

\newcommand\Vent{V_{S}}
\newcommand\comm[2]{{[}#1,#2{]}}
\newcommand\eqa{\begin{eqnarray}}
\newcommand\ena{\end{eqnarray}}

\newcommand\JR{j_{R}}
\newcommand\JL{j_{L}}

\newcommand\Tr{\mathrm{Tr}}
\newcommand{\me}{\mathrm{e}}

\newcommand\Yent{Y_{S}}

\newcommand\TR{T_{R}}
\newcommand\TL{T_{L}}

\newcommand\abs[1]{\left | #1 \right |}
\newcommand\junct{\mathit{junct}}
\newcommand\Qent{q_{S}}
\newcommand\Cent{{C_{S}}}
\newcommand\JRf{\tilde \jmath_{R}}
\newcommand\JLf{\tilde \jmath_{L}}
\newcommand\psiR{\psi_{R}}
\newcommand\psiL{\psi_{L}}
\newcommand\psiRf[1]{\tilde\psi_{R}(#1)}
\newcommand\psiLf[1]{\tilde\psi_{L}(#1)}
\newcommand\TRf[1]{\tilde T_{R}(#1)}
\newcommand\TLf[1]{\tilde T_{L}(#1)}
\begin{document}
%
%
\thispagestyle{empty}
\begin{center}
{\LARGE Entropy flow through near-critical quantum junctions
\footnotetext{In memory of Leo Kadanoff.}
\footnotetext{This is the second part of a two-part work originally published in 
2005 as  \cite{EntFlowI} and \cite{DFEntropyFlowII},
here revised for clarity following helpful suggestions of the referee.
The first part is \cite{EntFlowIJStatPhys} in this volume.
}}\\[6ex]
{\large Daniel Friedan}\\[1.5ex]
New High Energy Theory Center, Rutgers University, Piscataway, NJ, USA and \\\
Natural Science Institute, The University of Iceland, Reykjav\'\i k, Iceland\\
dfriedan@gmail.com\\[1.5ex]
February 21, 2017
\end{center}

\vspace*{2ex}

{\centering
\large\bfseries Abstract
\vskip1.5ex
}

This is the continuation of \cite{EntFlowIJStatPhys}.  Elementary 
formulas are derived for the flow of entropy through a 
circuit junction in a near-critical quantum circuit close to equilibrium, based 
on the structure of the energy-momentum tensor at the 
junction.  The entropic admittance of a near-critical 
junction in a bulk-critical circuit is expressed in terms of 
commutators of the chiral entropy currents.  The entropic 
admittance at low frequency, divided by the frequency, gives 
the change of the junction entropy with temperature --- the 
entropic ``capacitance''.  As an example, and as a check on 
the formalism, the entropic admittance is calculated 
explicitly for junctions in bulk-critical quantum Ising 
circuits (free fermions, massless in the bulk), in terms of 
the reflection matrix of the junction.  The half-bit of 
information capacity per end of critical Ising wire is 
re-derived by integrating the entropic ``capacitance'' with 
respect to temperature, from $T=0$ to $T=\infty$.

\setcounter{tocdepth}{2}
\tableofcontents
\newpage
%
%

\section{Summary}
This paper continues the elementary investigation of entropy 
flow in near-critical quantum circuits close to equilibrium begun in 
\onlinecite{EntFlowIJStatPhys}.  The entropy current operator, 
which is just the energy current divided by the temperature, 
$\Jent(x,t)=\kB \beta T_{t}^{x}(x,t)$, is used to analyze 
the flow of entropy through the general junction in a 
near-critical quantum circuit, especially when the quantum 
wires are critical in the bulk so that the only departure 
from criticality is in the junction.  Entropy flows in and 
out of the junction in response to changes in the entropic 
potentials in the wires, which are just the temperature 
drops in the wires.  The linear response coefficient for the 
entropy current is the entropic admittance of the junction.  
The junction entropy, $s(T)$, is the ``charge'' in the 
junction.  The entropic ``capacitance'' --- the temperature 
derivative, $\dif s/\dif T$ --- is extracted from the low 
frequency limit of the entropic admittance.  The information 
capacity of the junction, $s(\infty)-s(0)$ is found by 
integrating with respect to $T$.  The junction entropy 
itself needs a global calculation, but changes in the 
junction entropy can be determined locally, by studying the 
entropy currents near the junction.  A trivial example --- 
the general junction in a bulk-critical Ising circuit (i.e., 
free fermions, massless in the bulk) --- is done explicitly, 
to illustrate the formalism and check that it works.  
Definitions, notation and motivation are carried over from 
\onlinecite{EntFlowIJStatPhys}.  As explained there, the 
near-critical one-dimensional quantum systems under 
consideration are those that are described by 1+1 
dimensional relativistic quantum field theories,
and they are assumed to be close to equilibrium.

A junction consists of $N$ quantum wires joined at a single 
point.  The wires are labelled by indices $A,B=1\cdots N$.  
Each wire is parametrized by a spatial coordinate $x\ge 0$.  
The endpoints, $x=0$, are all identified to form a single 
point, the junction.  The entropy flow through the junction 
is analyzed in terms of the entropy currents in the $N$ 
external wires attached to the junction.  The junction is 
treated as a black box, from the point of view of the larger 
circuit containing it.  The single point $x=0$ might stand 
for a complicated sub-circuit, but the internal structure of 
the junction shows itself only by its effects on the flow of 
entropy in and out of the junction.

Suppose that a small variation is made in the entropic 
potential in each wire, $\Delta \Vent(t)^{B}=\me^{-i\omega 
t}\Delta \Vent(0)^{B}$, alternating at frequency $\omega$.  
The change in the entropic potential is the local 
temperature drop, $\Delta \Vent(t)^{B}=-\Delta T(t)^{B}$, on 
wire $B$ outside the junction.  The changing entropic 
potentials in the wires cause entropy currents
to flow through the wires, in and out of the 
junction, given by a linear response formula:
\eq
\Delta I_{S}(t)_{A} = \sum_{B=1}^{N}  \Yent(\omega)_{AB}\,
\Delta \Vent(t)^{B}
.
\label{eq:junction}
\en
The entropic admittance matrix, $\Yent(\omega)_{AB}$, 
describes the entropy flow characteristics of the junction.  
Entropic admittance has fundamental units $\kB ^{2}/\hbar$.

When the wires are bulk-critical, the entropy current is a 
sum of chiral currents, 
$\Jent(x,t)_{A}=\JR(x,t)_{A}-\JL(x,t)_{A}$, where 
$\JR(x,t)_{A}=\JR(x-vt)_{A}$ is the right-moving entropy 
current in wire $A$ and $\JL(x,t)_{A}=\JL(x+vt)_{A}$ is the 
left-moving entropy current, $v$ being the speed of 
``light'' in the relativistic quantum field theory of the 
one-dimensional bulk-critical system.  These are just the 
chiral energy currents of the conformally invariant 
bulk-critical system, divided by the temperature, $T$.  The 
Kubo formula for the entropic admittance is re-written in 
terms of the chiral entropy currents:
\eq
\Yent(\omega)_{AB} =
\frac{1}{i\omega} \int_{-\infty}^{\infty}\dif t_{1}\,\,
\me^{-i\omega (t_{1}-t_{2})} 
\expvalequil{
\frac{i}{\hbar}\comm{-\JL(0,t_{1})_{B}}
{ \JR(0,t_{2})_{A} -\JL(0,t_{2})_{A} }}
\label{eq:admittance}
.
\en
The time integral does not have the restriction $t_{1}\le 
t_{2}$ which expresses causality in the general Kubo formula:
the fact that the response happens after the perturbation.  
Causality is automatically enforced in 
(\ref{eq:admittance}) by chirality,
since a right-moving 
current cannot affect the junction, and a perturbation at 
the junction cannot produce any left-moving current.  
Equation (\ref{eq:admittance}) exhibits $\Yent(\omega)_{AB}$ 
as a well-behaved analytic function of $\omega$, free from 
the possibility of a short-time divergence,
which can arise in the general Kubo formula
in quantum field theory because of the 
restriction $t_{1}\le t_{2}$.

The Kubo formula is also re-written,
again for bulk-critical wires, in the form
\eq
\begin{split}
\frac{1}{i\omega}
&
\int_{-\infty}^{t_{2}} \dif t_{1} \,\,
\me^{-i\omega (t_{1}-t_{2})} 
\, \expvalequil{\frac{i}{\hbar}
\comm{\Jent(x_{1},t_{1})_{B}}{ \Jent(x_{2},t_{2})_{A} }}
=
\\
&
\qquad \me^{i\omega (x_{1}+x_{2})/v} \Yent(\omega)_{AB}
\\
&
-
\left (
\me^{i\omega \abs{x_{1}-x_{2}}/v} -\me^{i\omega (x_{1}+x_{2})/v} 
\right )
\delta_{AB}
\frac{c}{12} \frac{2\pi \kB ^{2}}{\hbar} 
\left [ 1 + 
\left ( \frac{\hbar \beta }{2\pi} \right )^{2} \omega^{2}
\right ] 
\label{eq:measure}
\end{split}
\en
where $x_{1},x_{2}>0$ are arbitrary points on the wires
and $c$ is the bulk conformal central 
charge.
This formula gives a way
to extract 
$\Yent(\omega)_{AB}$ from a measurement
performed at a distance
from the junction,
by generating entropy 
current at an arbitrary point $x_{1}>0$ in wire $A$, 
then detecting the induced entropy 
current at an arbitrary point $x_{2}>0$ in wire $B$.
These two elementary formulas for the entropic admittance
are the basic results of this paper.

The net entropy current leaving the junction is 
$\sum_{A=1}^{N}\Delta I_{S}(t)_{A}$, so the change in the 
junction entropy, $\Delta s(t)$, is given by 
$\partial_{t}\Delta s(t)=-i\omega \Delta s(t) = 
\sum_{A=1}^{N}\Delta I_{S}(t)_{A}$, so
\eq
\Delta s(t) = \frac1{i\omega} \sum_{A,B=1}^{N}
\Yent(\omega)_{AB}\,
\Delta \Vent(t)^{B}
.
\en
The change in the junction entropy with the junction 
temperature is
\eq
\frac{\dif s}{\dif T} = \lim_{\omega\rightarrow 0}
\; \frac1{i\omega} \sum_{A,B=1}^{N} \Yent(\omega)_{AB}
\label{eq:dsdT}
\en
because raising the temperature of the junction by $\Delta 
T$ is equivalent to dropping the temperature on all of the 
wires outside the junction by the same amount, $\Delta 
\Vent(t)^{B} = - \Delta T(t) ^{B}= \Delta T$.  
Equation (\ref{eq:dsdT}) also gives the specific heat of the 
junction, $T\dif s/\dif T$.  Integrating $\dif s/\dif T$ 
with respect to $T$ gives the junction entropy, $s(T)$, as a 
function of temperature, up to an additive constant.  
Integrating $\dif s/\dif T$ from $T=0$ to $T=\infty$ gives 
the total information capacity of the junction, 
$s(\infty)-s(0)$.  This does not give the junction entropy 
itself.  That requires a global calculation.  But 
\emph{changes} in $s(T)$ can be calculated locally, near the 
junction, using the entropy current.

$\Yent(\omega)_{AB}$ is a response function,
therefore analytic in the upper half-plane.
Equation (\ref{eq:admittance}) exhibits $\Yent(\omega)_{AB}$
as the Fourier transform in time of the 
equilibrium expectation value of a commutator of local 
operators,
so the only singularities in $\Yent(\omega)_{AB}$
at temperature $T>0$
should be poles in the lower half-plane.
The gap between the real axis and the nearest pole
of $\Yent(\omega)_{AB}$
is the inverse of the characteristic response time of the 
junction in the channel $AB$.
The zero temperature limit of
these gaps can serve as analogs
of the mass gap in a bulk
quantum field theory.

The conformal invariance of the bulk-critical wires implies
a vanishing formula:
\eq
\Yent(2\pi i/\hbar \beta)_{AB}  = 0
.
\en
In a separate paper,\cite{Friedan2006a} this vanishing formula is 
used to re-write the proof of the gradient formula for the 
junction beta-function in terms of the real time statistical 
mechanics of the entropy current.  The gradient formula, 
$\partial s/\partial \lambda^{a} 
=-g_{ab}(\lambda)\beta^{b}$, was originally proved in the 
less physical language of euclidean quantum field 
theory.\cite{FK}

As an example, and as a check on the formalism,
(\ref{eq:admittance}) is used to derive
an explicit formula
for $\Yent(\omega)_{AB}$ for the
general junction in a bulk-critical quantum Ising circuit.
The 1+1 dimensional quantum field theory of such a circuit
is a theory of free Majorana fermions,
massless in the bulk.\cite{AL1,GhZ,Chatterjee:1995sv,Konechny:2004dc}
An Ising junction is characterized by
its reflection matrix,
$R(\omega)_{B}^{A}$, which gives the amplitude
for left-moving fermions
of energy $\hbar \omega$ to enter the junction through 
wire $B$, then leave the junction as right-moving fermions
through wire $A$.
Equation (\ref{eq:admittance}) is used
to express the entropic admittance in 
terms of the reflection matrix:
\eqa
\Yent(\omega)_{AB}
&=&
\frac{\kB^{2} \hbar \beta^{2}}{2\pi\omega}
\int \int  \dif \omega _1 \dif \omega _2
\,\delta(\omega _1+\omega_2-\omega) \,
\left [
R(\omega_{1})^{A}_{B} R(\omega_{2})^{A}_{B}
-\delta^{A}_{B}
\right ]
\nonumber \\
&& \qquad\qquad\qquad
\frac18 (\omega_{1}-\omega_{2})^{2}
\left (
\frac1{1+\me^{-\beta\hbar \omega_{2}}}
- \frac1{\me^{\beta\hbar \omega_{1}} +1}
\right )
.
\ena
Substituting in (\ref{eq:dsdT})
gives the specific heat of the junction,
\eq
T\frac{\dif s}{\dif T} 
= 
\frac{k}{2\pi i} \int \dif \eta \;
\Tr \left [ R^{-1} R'(\eta)\right ]
\;
\frac{(\beta\hbar\eta)^{2}\me^{\beta \hbar \eta}}
{2 \left (\me^{\beta \hbar \eta}+1 \right )^{2}}
.
\label{eq:specificheat}
\en
Integrating with respect to $T$ gives $s(T)$ up to a constant.
These formulas apply to the general Ising junction.
$R(\omega)$ can depend on the
temperature, as when the junction has internal structure.
When $R(\omega)$ does \emph{not} depend on the temperature,
the integral over $T$ can be performed explicitly, giving
\eqa
s(T)-s(0) = 
\frac\kB{2\pi i} \int \dif \eta \;
\Tr \left [ R^{-1} R'(\eta) \right ]\;
\frac12
\left [
\frac{\beta \hbar \eta}{\me^{\beta \hbar \eta}+1 }
+\ln \left (
1+\me^{-\beta \hbar \eta}
\right )
\right ]
\label{eq:junctionentropy}
.
\ena
The information capacity is then
\eq
s(\infty)-s(0)
=
\frac\kB{2\pi i} \int \dif \eta \;
\Tr \left [ R^{-1} R'(\eta) \right ] \; \frac12 \ln 2
\label{eq:Isinginfo}
.
\en
The information capacity
consists of one half-bit for each pole of 
$R(\omega)^{A}_{B}$ in the lower half-plane
(or each zero in the upper half-plane),
when $R(\omega)$ is independent of temperature.

An elementary Ising junction, without
substructure, has a reflection matrix that is independent of 
temperature, of the form
\eq
R(\omega) = \frac{\omega-i\lambda^{T}\lambda}
{\omega+i\lambda^{T}\lambda}
\label{eq:elemR}
\en
where the matrix $\lambda_{B}^{A}$ parametrizes the junction 
couplings (generalizing the boundary magnetic field in a 
1-wire junction).  Equation (\ref{eq:Isinginfo}) then gives 
$s(\infty)-s(0)=(N/2)\kB \ln 2$.  The information capacity 
of an elementary Ising junction is one half-bit per end of 
critical Ising wire.  This reproduces a well-known result, 
at least for $N=1$.  For the case $N=1$, the formula for 
$s(T)-s(0)$, obtained from 
(\ref{eq:junctionentropy}) and~(\ref{eq:elemR}),
agrees with the formula for the boundary free energy and 
entropy of a single bulk-critical Ising wire, calculated 
directly.\cite{Chatterjee:1995sv,Konechny:2004dc} The case 
$N=2$ is equivalent to a pair of free Majorana fermion 
fields on a single wire with boundary.  When the reflection 
matrix is $U(1)$-invariant, this is equivalent to a single 
free massless Dirac fermion field on a wire with boundary, 
which is equivalent to the Toulouse limit of the spin-1/2 
Kondo model.  In this case, 
equation (\ref{eq:junctionentropy}) for the boundary entropy 
is identical to the formula for the boundary entropy in the 
spin-1/2 Kondo model, specialized to the Toulouse limit, as 
discussed in \onlinecite{TsvelickWiegmann:1983}, for 
example.  Equation (\ref{eq:specificheat}) for the junction 
specific heat is somewhat more general, allowing for the 
reflection matrix $R(\omega)$ to be temperature-dependent.  
Still, the field theory of a bulk-critical Ising circuit is 
a theory of free fermions, so there is no
difficulty to calculate the junction specific heat.  
It is done here as an 
illustration and check of the method of using the entropy 
current to describe the flow of entropy through the 
junction.  The direct calculation of the junction entropy is 
global.  First, the entropy of the whole system is 
calculated, then the bulk entropy of the wires is 
subtracted.  What remains is the junction entropy.  This is 
a subtle calculation, because boundary conditions are needed 
at the far ends of the wires, whose contributions to the 
total entropy must also be subtracted.  Using the entropy 
current to calculate \emph{changes} in the junction entropy 
is a local method that can be carried out in an arbitrarily 
small neighborhood of the junction.

\section{Entropy flow through a junction}
The energy-momentum tensor of a quantum circuit consists of 
a bulk part, located in the wires, and a contribution 
located in the junction (see 
Appendix~\ref{detail:boundaryenergy}).  The bulk 
energy-momentum tensor in wire $A$ is 
$T^{\mu}_{\nu}(x,t)_{A}$.  The junction contributes only to 
the energy density.
Its contribution is $\delta(x) T_{t}^{t}(t)_{\junct}$, where 
$T_{t}^{t}(t)_{\junct}$ is the junction energy.  The junction 
energy is conventionally written $T_{t}^{t}(t)_{\junct} = 
-\theta(t)$.  Energy and momentum are locally conserved in 
each bulk wire:
\eq
0=\partial_{\mu}T_{\nu}^{\mu}(x,t)_{A}
\:.
\en
Energy is conserved at the junction:
\eq
0 = -\partial_{t}\theta(t) + \sum_{A=1}^{N} 
T_{t}^{x}(0,t)_{A}
.
\label{eq:energy-inflow}
\en
The hamiltonian is
\eq
H_{0} =  -\theta(t) + \sum_{A=1}^{N}
\int_{0}^{\infty} \dif x \; T_{t}^{t}(x,t)_{A}
.
\en
The departure from criticality
is expressed by the trace of the energy-momentum tensor,
written $\Theta(x,t) = - T_{\mu}^{\mu}(x,t)$.
It consists of the bulk contribution in each wire,
$\Theta(x,t)_{A} = - 
T_{\mu}^{\mu}(x,t)_{A}$,
and the contribution located in the junction,
$\delta(x)\theta(t)$.
When the wires are bulk-critical,
the $\Theta(x,t)_{A}$ all vanish.
The departure from criticality 
is then entirely in the junction,
expressed by
\eq
\theta(t) = \beta^{a}(\lambda) \phi_{a}(t)
\en
where the $\phi_{a}(t)$ are the relevant 
and marginal operators
located in the junction,
and $\beta^{a}(\lambda)$
is the junction beta-function.
The junction is critical
when a change of scale has no effect,
which is equivalent to 
$\theta(t)$ being a multiple of the identity, 
which is equivalent to
\eq
0 = \partial_{t}\theta(t) = \sum_{A=1}^{N} T_{t}^{x}(0,t)_{A}
\en
which is also the condition that the boundary energy be
stationary.

The entropy current operators in the wires are
\eq
\Jent(x,t)_{A} = \kB  \beta T_{t}^{x}(x,t)_{A}
.
\en
The entropy density operators in the wires are
\eq
\rhoent(x,t)_{A} =
\kB  \beta T_{t}^{t}(x,t)_{A}
-\expvalequil{\kB  \beta T_{t}^{t}(x,t)_{A}}
.
\en
In addition, the junction
makes a contribution,
$\delta(x)\Qent (t)$,
to the entropy density operator,
where
\eq
\Qent (t) = - \kB  \beta \theta(t) 
+  \expvalequil{\kB  \beta \theta(t)}
.
\en
$\Qent (t)$ is
the junction entropy operator.
The entropy density operator measures the variation of 
the entropy density away from its equilibrium value.
$\Qent (t)$ measures the variation of the junction
entropy away from its
equilibrium value:
\eq
\Delta s(t) = \expval{\Qent (t)}
.
\en
Conservation of energy at the junction implies
conservation of entropy:
\eq
0= \partial_{t}\Qent(t)  + \sum_{A=1}^{N} \Jent(0,t)_{A}
.
\en
Any change in
the junction entropy
is equal to the net flow of entropy into the junction.

The entropy flow characteristics of the junction are its
responses to small changes in the entropic potentials
on the wires.
First consider a 1-wire junction,
which is simply the boundary of a wire.
Put a small alternating entropic potential,
$\Delta V_{S}(t)$, outside the junction,
constant along the wire:
\eq
\Delta \Phient(x,t) =
\Delta V_{S}(t) =  \me^{-i\omega t} \Delta V_{S}(0)
\qquad \mbox{for $x>x_{1}$}
\en
where $x_{1}$ is some
fixed point very close to $0$.
The hamiltonian is perturbed by
\eq
\Delta H = \int_{0}^{\infty} \dif x \,\,
\rhoent(x,t) \Delta \Phient(x,t)
= \Delta V_{S}(t) \int_{x_{1}}^{\infty} \dif x \,\,
\rhoent(x,t) 
.
\en
Entropy current will flow in response to the perturbation,
in and out of the 
boundary,
at frequency $\omega$.
In the linear response approximation, 
the induced current is
\eq
\Delta I_{S}(t)=
\Delta\expval{\Jent(0,t)} = \Yent(\omega) \Delta V_{S}(t)
,
\en
where the linear response coefficient, $\Yent(\omega)$, is 
the entropic admittance of the boundary.  $\Yent(0)=0$ for a 
1-wire junction, because the entropy flowing into the 
boundary encounters the same entropic potential as the 
entropy flowing out, when $\omega=0$.

A uniform small change in the entropic potential, constant 
over the whole system, has no effect, because the 
hamiltonian is perturbed by a multiple of itself.  
Therefore, raising the entropic potential by $\Delta 
V_{S}(t)$ everywhere outside the boundary is equivalent to 
lowering the potential on the boundary by $\Delta V_{S}(t)$.  
The perturbation can just as well be written $\Delta H(t) = 
-\Delta V_{S}(t) \Qent (t)$.  Now measure the entropy
current at a point $x_{2}>0$ very near the boundary.  The 
Kubo formula for the entropy current is
\eq
\begin{split}
\Delta\expval{\Jent(x_{2},t_{2})} &= \int_{-\infty}^{t_{2}} 
\dif t_{1} \,\, \expvalequil{\frac{i}{\hbar} \comm{\Delta 
H(t_{1})}{ \Jent(x_{2},t_{2}) }} \\
&= \Delta 
V_{S}(0,t_{2}) \int_{-\infty}^{t_{2}} \dif t_{1} \,\, 
\me^{-i\omega (t_{1}-t_{2})} \, 
\expvalequil{\frac{-i}{\hbar} \comm{\Qent (t_{1})}{ 
\Jent(x_{2},t_{2}) }}
\end{split}
\en
so the entropic admittance is
\eq
\Yent(\omega) = 
\lim_{x_{2}\rightarrow 0}
\int_{-\infty}^{t_{2}} \dif t_{1}
\,\, 
\me^{-i\omega (t_{1}-t_{2})} 
\, \expvalequil{\frac{-i}{\hbar}
\comm{\Qent (t_{1})}{ \Jent(x_{2},t_{2}) }}
\label{eq:YentQj}
\en
Conservation of entropy at the junction,
$0=\partial_{t}\Qent (t) +\Jent(0,t)$,
allows the Kubo formula to be written
\eq
\Yent(\omega) = 
\lim_{x_{2}\rightarrow 0}
\frac{1}{i\omega}
\int_{-\infty}^{t_{2}} \dif t_{1}
\,\, 
\me^{-i\omega (t_{1}-t_{2})} 
\, \expvalequil{\frac{i}{\hbar}
\comm{\Jent(0,t_{1})}{ \Jent(x_{2},t_{2}) }}
.
\en
The change of the boundary entropy is then
\eq
\Delta s(t) = \expval{\Qent (t)} =
(i\omega)^{-1}\expval{ \Jent (t)} =
(i\omega)^{-1}\Yent(\omega) \Delta V_{S}(t)
.
\en
In the static limit,
\eq
\Delta s = \Cent \Delta V_{S}
\en
where
\eq
\Cent =
\lim_{\omega\rightarrow 0}
\; (i\omega)^{-1}\Yent(\omega)
.
\en
Change of entropic charge
divided by change of entropic potential
might be called the entropic ``capacitance'' of the boundary
(abusing the electrical analogy).
The entropic potential of the boundary
has been \emph{lowered}
by $\Delta V_{S}$,
which is equivalent to \emph{raising} the temperature 
of the boundary by $\Delta T = \Delta V_{S}$,
so the entropic ``capacitance'' is
\eq
\Cent = \frac{\dif s \hfill}{\dif T \hfill}
.
\en
The specific heat of the 
boundary is $T\dif s/\dif T = T\Cent$.

For an $N$-wire junction,
impose small alternating potentials, $\Delta V_{S}(t)^{B}$,
$B=1\ldots N$,
on the wires outside the junction:
\eq
\Delta \Phient(x,t)^{B} =
\Delta V_{S}(t)^{B} =  \me^{-i\omega t} \Delta V_{S}(0)^{B} 
\qquad \mbox{for $x>x_{1}$}
.
\label{eq:perturbwires}
\en
The induced entropy current flowing out of the junction
through wire $A$ is
\eq
\Delta I_{S}(t)_{A}=
\lim_{x_{2}\rightarrow 0}\Delta\expval{\Jent(x_{2},t)_{A}}
= \sum_{B=1}^{N} \Yent(\omega)_{AB} \Delta V_{S}(t)^{B}
.
\en
The
linear response coefficients, $\Yent(\omega)_{AB}$,
form the entropic admittance matrix of the junction.
The entropic potential in the wires perturbs the hamiltonian 
by
\eq
\Delta H(t)_{1} = \sum_{B=1}^{N}
\int_{x_{1}}^{\infty} \dif x \; \rhoent(x,t)_{B}
\Delta V_{S}(t)^{B}
\en
which 
is gauge equivalent to
\eq
\Delta H(t) = \sum_{B=1}^{N}
\Jent(x_{1},t)_{B} \frac{1}{i\omega} \Delta V_{S}(t)^{B} 
.
\en
That is, the difference is
\eq
\Delta H(t)
-\Delta H(t)_{1}
=
\partial_{t}
\left [
\sum_{B=1}^{N}
\int_{x_{1}}^{\infty} \dif x \; \rhoent(x,t)_{B}
\frac{1}{i\omega}
\Delta V_{S}(t)^{B}
\right ]
.
\en
The Kubo formula for the induced entropy current
at a point, $x_{2}$, in wire $A$ is
\eq
\begin{split}
\Delta\expval{\Jent(x_{2},t_{2})_{A}} &=
\int_{-\infty}^{t_{2}} \dif t_{1}
\,\, 
\expvalequil{\frac{i}{\hbar} \comm{\Delta H(t_{1})_{1}}{
\Jent(x_{2},t_{2})_{A} }}
\\
&=
\int_{-\infty}^{t_{2}} \dif t_{1}
\,\, 
\expvalequil{\frac{i}{\hbar} \comm{\Delta H(t_{1})}{
\Jent(x_{2},t_{2})_{A} }}
\\
& \quad {} -
\frac{1}{i \omega}
\Delta V_{S}(t_{2})^{B}
\;
\expvalequil{
\int_{x_{1}}^{\infty} \dif x \; 
\frac{i}{\hbar} \comm{\rhoent(x,t_{2})_{B}}{ \Jent(x_{2},t_{2})_{A} }}
.
\label{eq:deltaj}
\end{split}
\en
If $x_{2}<x_{1}$, then the equal-time commutator in the second term is 
zero, because the operators are separated in space.
On the other hand, if $x_{2}>x_{1}$, then
\eqa
\expvalequil{
\int_{x_{1}}^{\infty} \dif x \; 
\frac{i}{\hbar} \comm{\rhoent(x,t_{2})_{B}}{ \Jent(x_{2},t_{2})_{A} }}
&=&
\kB \beta  
\expvalequil{
\int_{x_{1}}^{\infty} \dif x \; 
\frac{i}{\hbar} \comm{T_{t}^{t}(x,t_{2})_{B}}{ \Jent(x_{2},t_{2})_{A} }}
\nonumber \\
&=&
\kB \beta 
\delta_{AB}
\expvalequil{
\partial_{t}\Jent(x_{2},t_{2})_{A} }
\ena
which vanishes, since nothing changes with time
when the system is in equilibrium.
Since the second term on the rhs of (\ref{eq:deltaj})
always vanishes,
\eq
\begin{split}
\Delta\expval{\Jent(x_{2},t_{2})_{A}} 
&=
\sum_{B=1}^{N}
\int_{-\infty}^{t_{2}} \dif t_{1}
\\
&\qquad
\,\, 
\frac{1}{i\omega} \Delta V_{S}(t_{1})^{B}
\expvalequil{\frac{i}{\hbar} 
\comm{\Jent(x_{1},t_{1})_{B} }{\Jent(x_{2},t_{2})_{A} }}
\end{split}
\en
and the entropic admittance matrix is
\eq
\begin{split}
\Yent(\omega)_{AB} &= \lim_{x_{1},x_{2}\rightarrow 0}
\; \frac{1}{i\omega}
\int_{-\infty}^{t_{2}} \dif t_{1} \,\,
\\
&\qquad\qquad\qquad\qquad
\me^{-i\omega (t_{1}-t_{2})} 
\, \expvalequil{\frac{i}{\hbar}
\comm{\Jent(x_{1},t_{1})_{B}}{ \Jent(x_{2},t_{2})_{A} }}
.
\label{eq:junction-admittance}
\end{split}
\en
The change in the junction entropy is
\eq
\partial_{t}s(t) = 
\partial_{t}\expval{\Qent(t)}
= \sum_{A=1}^{N}
\expval{ -\Jent(0,t)_{A}}
\en
so
\eq
\Delta s(t) =
\frac{1}{i\omega}
\sum_{A=1}^{N} \sum_{B=1}^{N} 
\Yent(\omega)_{AB}
\Delta V_{S}(t)^{B}
.
\en
In the static limit,
\eq
\Delta s = \sum_{B=1}^{N}  C_{SB} \Delta V_{S}^{B}
\en
where
\eq
C_{SB} = \lim_{\omega\rightarrow 0}
\frac{1}{i\omega}
\sum_{A=1}^{N}
\Yent(\omega)_{AB}
.
\en
$C_{SB}$ is the entropic ``capacitance'' of the junction
(still abusing the electrical analogy).
Again, raising the temperature of the 
junction by $\Delta T$ is
equivalent to raising the entropic potential
everywhere outside the junction by 
$\Delta V_{S}^{B}=\Delta T$,
so
\eq
\frac{\dif s \hfill}{\dif T \hfill}
= C_{S} = \sum_{B=1}^{N} C_{SB} =
\lim_{\omega\rightarrow 0}
\frac{1}{i\omega}
\sum_{A,B=1}^{N} 
\Yent(\omega)_{AB}
\en
where $C_{S}$ is the total entropic ``capacitance'' of the junction.

\section{The continuity equation at a junction}

The change in the junction entropy is given by the time 
evolution equation
\eq
\partial_{t}\expval{\Qent(t)}
=
\expval{\partial_{t}\Qent(t)}
+
\expval{
\frac{i}\hbar \comm{\Delta H(t)}{\Qent(t)}
}
.
\en
The equal-time commutator vanishes, because
the entropic potential,
$\Phient(x,t)^{A} = \Delta \Vent(t)^{A}$,
is imposed slightly outside 
the junction, in the region $x>x_{1}$.
But the change in the junction entropy should properly include
also the changes in entropy in the wires very near the 
junction:
\eq
\partial_{t} 
\expval{
\sum_{A=1}^{N}
\int_{0}^{x_{2}}\dif x \,\,\rhoent(x,t)_{A}}
.
\en
This can be calculated using
the continuity equation for entropy flow in wires
derived in \onlinecite{EntFlowIJStatPhys}:
\eq
\begin{split}
\partial_{t} \expval{\rhoent(x,t)}
+\partial_{x}\expval{\Jent(x,t)}
&=
-\kB \beta \partial_{x}\Phient(x,t)\expval{\Jent(x,t)}
\\
&\qquad\qquad
- 
\kB \beta \partial_{x} 
\left [
\Phient(x,t)
\expval{  \Jent(x,t)}
\right ]
.
\end{split}
\en
The result is
\eq
\begin{split}
\partial_{t} 
\expval{\Qent(t)+\sum_{A=1}^{N}
\int_{0}^{x_{2}}\dif x \,\,\rhoent(x,t&)_{A}}
+\expval{\sum_{A=1}^{N}\Jent(x_{2},t)_{A}}
= 
\\
&
-\kB \beta
\Delta \Vent(t)^{A}
\expval{\Jent(x_{1},t)_{A}+\Jent(x_{2},t)_{A}}
.
\end{split}
\en
Note that this is still a local calculation at the junction.
The result depends only on the values of the entropic 
potential in the wires near the junction.
Now take $x_{1},x_{2}\rightarrow 0$,
including in the change of junction entropy
the changes of entropy in the wires near the junction:
\eq
\partial_{t}\expval{\Qent(t)}
+\sum_{A=1}^{N} \expval{\Jent(0,t)_{A}}
=
- 2 \kB \beta \sum_{A=1}^{N} \Delta\Vent(t)^{A}
\expval{  \Jent(0,t)_{A}}
.
\en
This is the continuity equation for entropy at the junction.
It is an exact equation.
There is no linear response approximation,
no assumption that $\Delta\Vent(t)^{A}$ is small,
so it is better written
\eq
\partial_{t}\expval{\Qent(t)}
+\sum_{A=1}^{N} \expval{\Jent(0,t)_{A}}
= - 2 \kB \beta \sum_{A=1}^{N} \Vent(t)^{A}
\expval{  \Jent(0,t)_{A}}
\en
or
\eq
\partial_{t}s(t)
+\sum_{A=1}^{N} I_{S}(t)_{A}
= - 2 \kB \beta \sum_{A=1}^{N} \Vent(t)^{A}
I_{S}(t)_{A}
\en
where $I_{S}(t)_{A}= \expval{  \Jent(0,t)_{A}}$.

\section{Junctions in bulk-critical quantum circuits}

Assume now that all the wires are critical in the bulk:
\eq
\Theta(x,t)_{A} = - T_{\mu}^{\mu}(x,t)_{A}=0
.
\en
Any departure from criticality is in the junction.
The entropy current in each bulk-critical wire
separates into chiral entropy currents
(see Appendix~\ref{app:chiralcommutators}):
\eqa
\Jent(x,t)_{A} &=& \JR(x,t)_{A} - \JL(x,t)_{A} \\
\rhoent(x,t)_{A} &=& v^{-1} \JR(x,t)_{A} + v^{-1}\JL(x,t)_{A}  
\ena
where
\eq
\JR(x,t)_{A} = \JR(x-vt)_{A}
\en
is the right-moving entropy current, and
\eq
\JL(x,t)_{A} = \JL(x+vt)_{A}
\en
is the left-moving entropy current.
The operators $\JR(x-vt)_{A}$ and $\JL(x+vt)_{A}$ are 
defined on the entire real line, because, although $x$
cannot be negative, the time 
$t$ can range from $-\infty$ to $+\infty$.
The rate of change of the junction entropy is
the net rate of inflow,
\eq
\partial_{t}\Qent  = \sum_{A=1}^{N}
 \JL(0,t)_{A} - \sum_{A=1}^{N} \JR(0,t)_{A}
.
\label{eq:chiralinflow}
\en

The equilibrium expectation values of commutators of the 
chiral currents vanish in regions of space-time where 
effects are forbidden by causality and chirality:
\eqa
\expvalequil{
\comm{\JR(x_{1},t_{1})_{B}}{ \JR(x_{2},t_{2})_{A} }}
&=& 0
\quad \mbox{if $A\ne B$ or $t_{2}-t_{1} \ne (x_{2}-x_{1})/v$}
\label{eq:identityone}
\\
\expvalequil{
\comm{\JL(x_{1},t_{1})_{B}}{ \JL(x_{2},t_{2})_{A} }}
&=& 0
\quad \mbox{if $A\ne B$ or $t_{2}-t_{1} \ne (x_{1}-x_{2})/v$}
\label{eq:identitytwo}
\\
\expvalequil{
\comm{\JR(x_{1},t_{1})_{B}}{ \JL(x_{2},t_{2})_{A} }}
&=& 0
\quad \mbox{if $t_{1}-t_{2}< (x_{1}+x_{2})/v$}
\label{eq:identitythree}
\\
\expvalequil{
\comm{\JL(x_{1},t_{1})_{B}}{ \JR(x_{2},t_{2})_{A} }}
&=& 0
\quad \mbox{if $t_{2}-t_{1}<(x_{1}+x_{2})/v$}
.
\label{eq:identityfour}
\ena
The first two identities express the fact that any 
commutator of two currents of the same chirality is an 
equal-time commutator.  The last two identities express the 
fact that the left-moving current at $x_{1}$ cannot affect 
the right-moving current at $x_{2}$ before a signal has time 
to travel from $x_{1}$ to the junction then back to $x_{2}$, 
and the fact that the right-moving current cannot affect the 
left-moving current at all.
For the first identity, use time-translation invariance and 
chirality to write
\eq
\begin{split}
\expvalequil{
\comm{\JR(x_{1},t_{1})_{B}}{ \JR(x_{2},t_{2})_{A} }}
= \qquad\qquad\qquad\qquad\qquad\qquad
\\
\expvalequil{
\comm{\JR(x_{1}-vt_{1}+vt_{2}+vt,t_{2})_{B}}{ \JR(x_{2}+vt,t_{2})_{A} }}
\label{eq:JRJRasequaltime}
\end{split}
\en
for any $t$ sufficiently large, such that $x_{2}+ v t>0$ and 
$x_{1}-vt_{1}+vt_{2}+vt>0$.  The commutator on the rhs is 
then an equal-time commutator, which vanishes if $A\ne B$ or 
if $x_{1}-x_{2}-vt_{1}+vt_{2} \ne 0$.  The second identity, 
(\ref{eq:identitytwo}), is derived in the same way.
For the third identity,
(\ref{eq:identitythree}),
write
\eq
\begin{split}
\expvalequil{
\comm{\JR(x_{1},t_{1})_{B}}{ \JL(x_{2},t_{2})_{A} }}
= \qquad\qquad\qquad\qquad\qquad\qquad
\\
\expvalequil{
\comm{\JR(x_{1}-vt_{1}+vt_{2}-vt,t_{2})_{B}}{ \JL(x_{2}+vt,t_{2})_{A} }}
\end{split}
\en
whenever $x_{1}-vt_{1}+vt_{2} - vt>0$ and $x_{2}+vt>0$,
which is possible if and only if $t_{1}-t_{2}< (x_{2}+x_{1})/v$.
The commutator on the rhs is then an equal-time commutator
at nonzero spatial separation, so is zero.
The fourth identity, (\ref{eq:identityfour}),
follows in the same way.

For $t_{1}<t_{2}$, equations (\ref{eq:identityone}),
(\ref{eq:identitytwo}), and~(\ref{eq:identitythree})
imply that
\eq
\begin{split}
\expvalequil{
\comm{\Jent(x_{1},t_{1})_{B}}{ \Jent(x_{2},t_{2})_{A} }}
=\qquad\qquad\qquad\qquad\qquad\qquad\qquad\qquad
\\
\left \{
\begin{array}{ll}
\expvalequil{\comm{\JR(x_{1},t_{1})_{B}-\JL(x_{1},t_{1})_{B}}{
\JR(x_{2},t_{2})_{A} }}
& \mbox{\quad if $x_{1}<x_{2}$}
\\[1ex]
\expvalequil{\comm{-\JL(x_{1},t_{1})_{B}}{ 
\JR(x_{2},t_{2})_{A} -\JL(x_{2},t_{2})_{A} }}
& \mbox{\quad if $x_{1}>x_{2}$.}
\end{array}
\right .
\end{split}
\en
Then
\eq
\begin{split}
\frac{1}{i\omega}
&\int_{-\infty}^{t_{2}} \dif t_{1} \,\,
\me^{-i\omega (t_{1}-t_{2})} 
 \, \expvalequil{\frac{i}{\hbar}
\comm{\Jent(x_{1},t_{1})_{B}}{ \Jent(x_{2},t_{2})_{A} }}
=
\\
&\frac1{i \omega} \int_{-\infty}^{\infty}\dif t_{1}\,\,
\me^{-i\omega (t_{1}-t_{2})} 
\qquad\qquad\qquad\qquad\qquad
\label{eq:JSJSanywhere}
\\
\qquad
\qquad
&\left \{
\begin{array}{ll}
\expvalequil{\frac{i}{\hbar}
\comm{\JR(x_{1},t_{1})_{B}-\JL(x_{1},t_{1})_{B}}{ \JR(x_{2},t_{2})_{A} }}
& \mbox{if $x_{1}<x_{2}$} \\[1.5ex]
\expvalequil{\frac{i}{\hbar}
\comm{-\JL(x_{1},t_{1})_{B}}{ \JR(x_{2},t_{2})_{A} -\JL(x_{2},t_{2})_{A} }}
& \mbox{if $x_{1}>x_{2}$}
\end{array}
\right .
\end{split}
\en
where the restriction to $t_{1}\le t_{2}$ in the time
integrals on the rhs has been dropped
because the integrands vanish when $t_{1} > t_{2}$,
by 
(\ref{eq:identityone}),
(\ref{eq:identitytwo}), and~(\ref{eq:identityfour}).

A possible ambiguity in
equation (\ref{eq:junction-admittance}) for $\Yent(\omega)_{AB}$
is now apparent.
Taking the limit $x_{1},x_{2}\rightarrow 0$ with $x_{1}<x_{2}$
gives one expression for $\Yent(\omega)_{AB}$:
\eq
\begin{split}
\Yent(\omega)_{AB}^{12} =
\frac1{i \omega} \int_{-\infty}^{\infty}\dif t_{1}\,\,
\me^{-i\omega (t_{1}-t_{2})} 
\qquad\qquad\qquad\qquad\qquad\qquad
\\
\expvalequil{\frac{i}{\hbar}
\comm{\JR(0,t_{1})_{B}-\JL(0,t_{1})_{B}}{ \JR(0,t_{2})_{A} }}
.
\label{eq:YentchiralA}
\end{split}
\en
Taking the limit with $x_{1}>x_{2}$ gives another expression:
\eq
\begin{split}
\Yent(\omega)_{AB}^{21} =
\frac1{i \omega} \int_{-\infty}^{\infty}\dif t_{1}\,\,
\me^{-i\omega (t_{1}-t_{2})} 
\qquad\qquad\qquad\qquad\qquad\qquad
\\
\expvalequil{\frac{i}{\hbar}
\comm{-\JL(0,t_{1})_{B}}{ \JR(0,t_{2})_{A} -\JL(0,t_{2})_{A} }}
.
\label{eq:YentchiralB}
\end{split}
\en
The difference involves only
the commutators of currents of the same chirality.
These commutators are 
completely determined by bulk conformal invariance (see 
(\ref{eq:TRTRcomm}) and~(\ref{eq:TLTLcomm}) of
Appendix~\ref{app:chiralcommutators}), with the result that
\eq
\Yent(\omega)_{AB}^{21} - \Yent(\omega)_{AB}^{12} 
=
\delta_{AB}
2 \kB \beta \expvalequil{\Jent(0,t)_{A} }
,
\en
so the ambiguity only exists when there are nonzero entropy 
currents flowing in equilibrium.  
As explained in \cite{EntFlowIJStatPhys}
at the end of section 8,
persistent equilibrium currents
are possible because
of the isolation of the near-critical degrees of freedom
from the environment.
In the circuit laws for 
entropy, the choice between $\Yent(\omega)_{AB}^{12}$ and 
$\Yent(\omega)_{AB}^{21}$ is a matter of convention, as long 
as the convention is used consistently.  For the local 
analysis of a junction connected to open wires, the 
appropriate choice depends on the situation.  To describe 
the change in the entropy currents due to a change $\Delta 
T(t)$ in the junction temperature, the appropriate choice is 
$\Yent(\omega)_{AB}^{12}$:
\eq
\Delta I_{S}(t)_{A}
= \sum_{B=1}^{N} \Yent(\omega)_{AB}^{12} \Delta T(t)
.
\label{eq:DeltaI}
\en
The entropic potential on the junction is changed by 
$-\Delta T(t)$, then the resulting currents are observed 
outside the junction, i.e., $x_{1}<x_{2}$.  On the other 
hand,
to describe the 
change in the junction entropy due to changes, $\Delta 
T(t)^{B}= -\Delta V_{S}(t)^{B}$, in the temperatures of the 
wires, the appropriate choice is $\Yent(\omega)_{AB}^{21}$:
\eq
\Delta s(t) =
\sum_{B=1}^{N}
\left (
\frac{1}{i\omega}
\sum_{A=1}^{N}  
\Yent(\omega)_{AB}^{21}
\right )
\Delta V_{S}(t)^{B}
.
\label{eq:Deltas}
\en
The entropic potential is changed in the wires outside the 
junction, then the net flow of entropy into the junction 
in measured at the junction, i.e., $x_{1}>x_{2}$.

For either choice of convention, $\Yent(\omega)_{AB}$ is 
manifestly a well-defined analytic function of the 
frequency, given by (\ref{eq:YentchiralA})
or~(\ref{eq:YentchiralB}).  The original Kubo formula, 
(\ref{eq:junction-admittance}), has the possibility 
of a short-time divergence, because of the restriction to 
$t_{1}\le t_{2}$ in the integral over time.  There is no 
such possibility in (\ref{eq:YentchiralA})
or~(\ref{eq:YentchiralB}), since $\Yent(\omega)_{AB}$ is 
simply the Fourier transform in time of a commutator of 
local operators.

Define the Fourier modes of the chiral entropy currents by
\eqa
\JR(x,t)_{A}
&=& \frac1{2\pi} \int_{-\infty}^{\infty} \dif \omega \,\, 
\me^{-i\omega (t-x/v)} \, \JRf(\omega)_{A}
\\[1ex]
\JL (x,t)_{A}
&=&  \frac1{2\pi} \int_{-\infty}^{\infty} \dif \omega  \,\, 
\me^{-i\omega (t+x/v)} \, \JLf(\omega)_{A}
.
\ena
Equations~(\ref{eq:YentchiralA}) and~(\ref{eq:YentchiralB})
are equivalent to
\eqa
2\pi \delta (\omega'+\omega) \,\hbar \omega 
\Yent(\omega)_{AB}^{12}
&=&
\expvalequil{\comm{\JRf(\omega')_{B}-\JLf(\omega')_{B}}{ 
\JRf(\omega)_{A} }}
\label{eq:YentmodesA}
\\
2\pi \delta (\omega'+\omega) \,\hbar \omega 
\Yent(\omega)_{AB}^{21}
&=&
\expvalequil{\comm{-\JLf(\omega')_{B}}{
\JRf(\omega)_{A} -\JLf(\omega)_{A} }}
.
\label{eq:YentmodesB}
\ena
These formulas are used below to calculate
$\Yent(\omega)_{AB}$ for junctions in bulk-critical Ising 
circuits.

From now on, to simplify the discussion,
assume that no equilibrium entropy currents 
are flowing through the junction.
With this assumption,
\eq
\Yent(\omega)_{AB} = \Yent(\omega)_{AB}^{12}
= \Yent(\omega)_{AB}^{21}
.
\en
When there are no equilibrium entropy currents,
bulk conformal invariance
implies (see (\ref{eq:TTnocurrent}) of
Appendix~\ref{app:chiralcommutators}):
\eqa
\expvalequil{\comm{\JRf(\omega')_{A}}{\JRf(\omega)_{B}}}
&=&
- \delta_{AB} \; 2\pi\, \delta(\omega'+\omega)
\frac{c}{12} 2\pi \kB^{2}   \omega
\left [
1+ \left ( \frac{\hbar \beta}{2\pi}
\right )^{2}
\omega^{2}
\right ] 
\\
\expvalequil{\comm{\JLf(\omega')_{A}}{\JLf(\omega)_{B}}}
&=&
- \delta_{AB} \; 2\pi\, \delta(\omega'+\omega)
\frac{c}{12} 2\pi \kB^{2}   \omega
\left [
1+ \left ( \frac{\hbar \beta}{2\pi}
\right )^{2}
\omega^{2}
\right ] 
\ena
This allows (\ref{eq:JSJSanywhere}) to be re-written:
\eq
\begin{split}
&
\frac{1}{i \omega}
\int_{-\infty}^{t_{2}} \dif t_{1} \,\,
\me^{-i\omega (t_{1}-t_{2})} 
\, \expvalequil{\frac{i}{\hbar}
\comm{\Jent(x_{1},t_{1})_{B}}{ \Jent(x_{2},t_{2})_{A} }}
=
\\
&\qquad\qquad \me^{i\omega (x_{1}+x_{2})/v} \Yent(\omega)_{AB}
\\
&
- \left (
\me^{i\omega \abs{x_{1}-x_{2}}/v} -\me^{i\omega (x_{1}+x_{2})/v} 
\right )
\delta_{AB}
\frac{c}{12} \frac{2\pi \kB ^{2}}{\hbar} \left [ 1 + 
\left ( \frac{\hbar \beta }{2\pi} \right )^{2} \omega^{2}
\right ] 
\label{eq:Yentdistance}
\end{split}
\en
for any $x_{1},x_{2}\ge 0$.
The lhs of this equation describes an experiment in which 
the system is perturbed at some arbitrary point, $x_{1}$,
away from the junction,
then
the response is detected at a second
arbitrary point, $x_{2}$, also away from the junction.
Equation (\ref{eq:Yentdistance}) says that
the entropic admittance of the junction
can be extracted from this measurement.

\section{Properties of $\Yent(\omega)_{AB}$
in a bulk-critical quantum circuit}

The entropic admittance of a junction
in a bulk-critical circuit
satisfies:
\begin{quote}
\begin{enumerate}[itemsep=10pt]
\makeatletter
\renewcommand{\theenumi}{\Alph{enumi}}
\renewcommand{\labelenumi}{\theenumi.}
\setlength\labelsep{2ex}
\makeatother
\item \label{propAnal}
$\Yent(\omega)_{AB}$ is analytic 
in the upper half-plane.
\item \label{propReal}
$\overline{\Yent(\omega)}_{AB} = \Yent(-\bar\omega)_{AB}$.
\item \label{propPos}
$\Yent(\omega) + \Yent(\omega)^{\dagger} \le 0$
for all real $\omega$.
\item \label{propSum}
$\sum_{A=1}^{N} \Yent(0)_{AB} =\sum_{B=1}^{N} \Yent(0)_{AB} = 0$.
\item \label{propMero}
The only singularities of 
$\Yent(\omega)_{AB}$ are poles lying below the real axis.
\item \label{propConf}
$\Yent(2\pi i/\hbar\beta)_{AB} = 0$.
\end{enumerate}
\end{quote}
All of these properties can be 
checked in the explicit example
given by
(\ref{eq:IsingYent1wire}) below.

Property~\ref{propAnal} expresses causality.
The reality condition, property~\ref{propReal},
follows directly from
(\ref{eq:YentmodesA}) and~(\ref{eq:YentmodesB})
and the fact that the currents are self-adjoint.
For the positivity condition, property~\ref{propPos},
write (\ref{eq:YentmodesA}) and~(\ref{eq:YentmodesB})
in terms of the 
equilibrium two-point 
functions:
\eqa
2\pi \delta (\omega'+\omega) \,
\frac{\hbar \omega \Yent(\omega)_{AB}^{12}}
{1-\me^{\beta \hbar \omega}}
&=&
\expvalequil{\JRf(\omega')_{B}\,
\JRf(\omega)_{A} }
-
\expvalequil{\JLf(\omega')_{B}\,
\JRf(\omega)_{A} }
\label{eq:YeqA}
\\[1ex]
2\pi \delta (\omega'+\omega) \,
\frac{\hbar \omega \Yent(\omega)_{AB}^{21}}
{1-\me^{\beta \hbar \omega}}
&=&
\expvalequil{\JLf(\omega')_{B}\,
\JLf(\omega)_{A} }
-\expvalequil{\JLf(\omega')_{B}\,
\JRf(\omega)_{A}  }
.
\label{eq:YeqB}
\ena
Take the complex conjugate of (\ref{eq:YeqB}),
use the self-adjointness of the currents and
the
the reality condition, property~\ref{propReal},
and exchange $A$ with $B$ and $\omega$ with $\omega'$
to get
\eq
2\pi \delta (\omega'+\omega) \,
\frac{\hbar \omega \overline{\Yent(\omega)_{BA}^{21}}}
{1-\me^{\beta \hbar \omega}}
=
\expvalequil{\JLf(\omega')_{B}\,
\JLf(\omega)_{A} }
-
\expvalequil{\JRf(\omega')_{B}\,
\JLf(\omega)_{A} }
\label{eq:YeqBcc}
\en
Add (\ref{eq:YeqA}) and~(\ref{eq:YeqBcc})
and use the self-adjointness of the currents to get
property~\ref{propPos}
(still assuming no equilibrium entropy currents,
so $\Yent(\omega)_{AB}^{21}=\Yent(\omega)_{AB}^{12}=
\Yent(\omega)_{AB}$).

For property~\ref{propSum},
note that conservation at the junction,
(\ref{eq:chiralinflow}),
implies
\eq
0 = \sum_{A=1}^{N} \left [ \JLf(0)_{A}
- \JRf(0)_{A} \right ]
.
\en
In (\ref{eq:YeqA}), set $\omega'=0$ to get
\eq
2\pi \delta (\omega) \,
\left ( \frac{-1}{\beta} \right )
\sum_{B=1}^{N}
\Yent(0)_{AB}^{12}
=
\expvalequil{\sum_{B=1}^{N}
\left [ \JRf(0)_{B} - \JLf(0)_{B} \right ]
\JRf(\omega)_{A} }
= 0
\en
so $\sum_{B=1}^{N}
\Yent(0)_{AB}^{12}=0$.
In (\ref{eq:YeqB}), set $\omega=0$ to get
\eq
2\pi \delta (\omega') \,
\left ( \frac{-1}{\beta} \right )
\sum_{A=1}^{N}
\Yent(0)_{AB}^{21}
=
\expvalequil{\JLf(\omega')_{B}\,
\sum_{A=1}^{N}
\left [ \JLf(0)_{A} - \JRf(0)_{A} \right ]
}
= 0
\en
so $\sum_{A=1}^{N}
\Yent(0)_{AB}^{21} = 0$.
These results are exactly what is required
in order that equations (\ref{eq:DeltaI})
and~\ref{eq:Deltas} be physically sensible at $\omega=0$.

Property~\ref{propMero} follows from 
(\ref{eq:YentchiralA}) or~(\ref{eq:YentchiralB}), each 
of which displays $\Yent(\omega)_{AB}$ as the Fourier 
transform in time of the equilibrium expectation value of a 
commutator of local operators.  The junction is a bounded 
system at nonzero temperature, so such an expectation value
is a meromorphic function of $\omega$.
Causality implies that all its
singularities lie below the real axis.

Property~\ref{propConf}, $\Yent(2\pi i/\hbar\beta)_{AB} = 0$,
is derived from
bulk conformal invariance,
by Wick rotating to euclidean time $\tau=it$.
The euclidean space-time of each bulk wire is a semi-infinite 
cylinder: $x > 0$, $\tau \sim \tau + \hbar \beta$.
This euclidean space-time can be reinterpreted with
$x/v$ as the euclidean time coordinate and
and $v \tau$ as the spatial 
coordinate.
Space is then a circle of circumference $\hbar v\beta $. 
In this quantization,
correlation functions on the semi-infinite cylinder are 
expectation values between
the conformally invariant ground state
at late time, at large $x/v$,
and a boundary state at time $x/v=0$.

In this quantization,
\eq
\TR(x+i v \tau)_{A}
=
\frac{-\hbar v^{2}}{2\pi}
\sum_{n=-\infty}^{\infty}
\me^{-2 \pi n (x+ i v \tau)/\hbar v  \beta}
L_{n}
\en
where the $L_{n}$ are the Virsaro operators of the bulk 
conformal field theory in wire $A$
(the constant of proportionality is 
found in Appendix~\ref{detail:boundaryenergy}).
The $L_{n}$ for
$n\le 1$,
annihilate the conformally invariant ground 
state when they
act to the left, at large time $x/v$.
In a correlation function containing
$\TR(x+i v \tau)_{A}$
for sufficiently large $x$,
the contributions of the operators
$L_{n}$, $n\le 1$, vanish,
because they are acting
on the conformally invariant ground 
state.
The two-point function
given by the Fourier transform of
(\ref{eq:YeqA}) is:
\eq
\begin{split}
(\kB\beta)^{2}
\expvalequil{
\left [ \TR(0)_{B} - \TL(0)_{B} \right ]
\;
\TR(x-vt)_{A}}
=\qquad\qquad\qquad
\\
\frac1{2\pi}
\int \dif \omega  \,\, 
\me^{-i\omega (t-x/v)}
\frac{\hbar \omega \Yent(\omega)_{AB}}
{1-\me^{\beta \hbar \omega}}
.
\end{split}
\en
The product of operators on the lhs is time-ordered for 
$t<0$ (in the original, physical quantization).
The Wick rotatation of this equation is sensible in
the region $ -\hbar\beta < \tau < 0$,
\eq
\begin{split}
(\kB\beta)^{2}
\expvalequil{
\left [ \TR(0)_{B} - \TL(0)_{B} \right ]
\,
\TR(x+iv\tau)_{A}}
=\qquad\qquad\qquad
\\
\frac1{2\pi}
\int \dif \omega  \,\, 
\me^{-\omega(\tau + i x/v)}
\frac{\hbar \omega \Yent(\omega)_{AB}}
{1-\me^{\beta \hbar \omega}}
,
\end{split}
\en
because the integrand decays exponentially when $\omega 
\rightarrow \pm\infty$,
as long as $ -\hbar\beta < \tau < 0$.
Deforming the integration contour into the upper half-plane 
yields a sum over the thermal poles:
\eq
\begin{split}
(\kB\beta)^{2}
\expvalequil{
\left [ \TR(0)_{B} - \TL(0)_{B} \right ]
\;
\TR(x+iv\tau)_{A}}
=\qquad\qquad\qquad
\\
\sum_{n=1}^{\infty}
\frac{2\pi n}{\beta}
\Yent(2\pi i n/\hbar\beta)_{AB}
\;
\me^{-2\pi n(x+iv\tau)/\hbar v \beta}
\end{split}
\en
The $n=1$ term must vanish because
it is due to the conformal generator $L_{1}$
acting on the conformally invariant ground state.
So bulk conformal invariance implies 
$\Yent(2\pi i /\hbar\beta)_{AB} =0$.

\section{``Mass gaps'' for junctions}
The real-time response functions
\eq
\frac1{2\pi} \int \dif \omega  \,\, \me^{-i\omega (t_{2}-t_{1})}
\Yent(\omega )_{AB}
\en
describe the entropy flowing out of the junction at time 
$t_{2}$ in wire $A$, in response to a perturbation in wire 
$B$ at time $t_{1}<t_{2}$.  The integral can be evaluated by 
deforming the contour of integration into the lower 
half-plane, since $t_{2}-t_{1}>0$.  The response decays 
as $\me^{-t/t_{\mathit{r}}}$, where 
$1/t_{\mathit{r}}$ is the gap between the real $\omega$-axis 
and the nearest singularity in the lower half-plane.  The 
time $t_{\mathit{r}}$ is the the characteristic response 
time of the junction in the channel $AB$.  At nonzero 
temperature, there will always be a gap.  In the 
Ising example, (\ref{eq:IsingYent1wire}) below, the 
characteristic response time of a 1-wire junction is 
$t_{\mathit{r}}=\lambda^{2}+\pi/\hbar\beta$, where $\lambda$ 
is the boundary coupling constant (the boundary magnetic 
field).

The interesting question is,
what happens to these gaps in the limit $T \rightarrow 0$?
The question is better posed in terms of the energy 
response functions
\eqa
\expval{\frac{i}\hbar \comm{
T_{t}^{x}(0,t_{1})_{B}}{ T_{t}^{x}(0,t_{2})_{A}
}}
\ena
because of the extra factor $1/T$ in the entropy current.
When the junction is non-critical,
then some of the $T=0$ response functions
will decay exponentially in time.
If the junction is critical, then the $T=0$ response functions 
will decay as powers of the time, or else vanish identically.
The gaps, $1/t_{\mathit{r}}$, in each channel
might be interpreted as the ``mass gaps''
of the junction,
analogous to the mass gap in a bulk system.

\section{Example: Junctions in bulk-critical Ising circuits}

As an example, consider the general junction in a 
bulk-critical Ising circuit.  The wires are in the 
universality class of the 1+1 dimensional Ising model at its 
critical point.  The circuit is described by the quantum 
field theory of a free Majorana fermion, massless in the 
bulk.  There are two real, chiral, free 
fermion fields in each wire:
\eqa
\psiR(x,t)^{A} &=& \psiR(x-vt)^{A} \\
\psiL(x,t)^{A} &=& \psiL(x+vt)^{A}
.
\ena
The operators $\psiR(x-vt)^{A}$ and $\psiL(x+vt)^{A}$ are 
defined on the entire real line, because, although $x$
cannot be negative, the time 
$t$ can range from $-\infty$ to $+\infty$.
The fermion fields
satisfy canonical equal-time anti-commutation relations
in the bulk.
For $x,x'>0$:
\eqa
\comm{\psiR(x',t)^{A}}{\psiR(x,t)^{B}}_{+} &=&
\delta^{AB} \; \hbar v \, \delta(x'-x) \\
\comm{\psiL(x',t)^{A}}{\psiL(x,t)^{B}}_{+} &=&
\delta^{AB} \; \hbar v \, \delta(x'-x)\\
\comm{\psiR(x',t)^{A}}{\psiL(x,t)^{B}}_{+} &=& 0
.
\label{eq:psiRL}
\ena
Their Fourier transforms are
\eqa
\psiR(x-v t)^{A} &=& 
\frac1{2\pi}\
\int \dif \omega  \,\,\me^{-i \omega (t-x/v)} \psiRf{\omega }^{A}
\\
\psiL(x+v t)^{A} &=& 
\frac1{2\pi}\int\dif \omega \,\,\me^{-i \omega (t+x/v)} 
\psiLf {\omega}^{A} 
\ena
The hamiltonian, $H_{0}$, acts by
\eqa
\comm{H_{0}}{ \psiRf {\omega }^{A}} &=& -\hbar\omega 
\psiRf {\omega }^{A} \\
\comm{H_{0}}{ \psiLf {\omega }^{A}} &=& -\hbar\omega 
\psiLf {\omega }^{A}
.
\ena
The junction dynamics is encoded in the reflection matrix,
$R(\omega)^{A}_{B}$:
\eq
\psiRf {\omega }^{A}= \sum_{B=1}^{N}
R(\omega)^{A}_{B} \psiLf {\omega}^{B}
\label{eq:reflection}
\en
or, suppressing indices,
\eq
\psiRf {\omega }= R(\omega) \psiLf {\omega}
.
\en
The Fourier modes satisfy the anti-commutation relations
\eqa
\comm{\psiRf {\omega' }^{A}}{ \psiRf {\omega }^{B}}_{+} &=&  
\delta^{AB} \, 2\pi \hbar\, \delta(\omega' +\omega )
\label{eq:psiRpsiR}
\\
\comm{\psiLf {\omega' }^{A}}{ \psiLf {\omega }^{B}}_{+} &=& 
\delta^{AB}\,  2\pi \hbar\, \delta(\omega' +\omega )
\\
\comm{\psiLf {\omega' }^{A}}{ \psiRf {\omega }^{B}}_{+} &=& 
\sum_{B'=1}^{N}\delta^{AB'}R(\omega)^{B}_{B'}\,
 2\pi \hbar\, \delta(\omega' +\omega )
.
\label{eq:psiLpsiR}
\ena
The matrix $R(\omega)$ is analytic in the upper 
half-plane, 
by (\ref{eq:psiRL}), and satisfies
the reality and unitarity conditions
\eq
R(-\omega) = \overline{R(\omega)}
\en
\eq
R(-\omega)^{T} R(\omega) = 1
.
\en
The anti-commutation relations and the action of the 
hamiltonian determine the equilibrium two-point functions:
\eqa
\expvalequil{\psiRf {\omega' }^{A}
\psiRf {\omega }^{B}}  &=&
(1+ \me^{\beta \hbar\omega })^{-1}
\delta^{AB}\,
2\pi \hbar\delta(\omega' +\omega)
\label{eq:twopointequilRR}
\\
\expvalequil{\psiLf {\omega' }^{A}
\psiLf {\omega }^{B}}  &=&
(1+ \me^{\beta \hbar\omega })^{-1}
\delta^{AB}\,
2\pi \hbar\delta(\omega' +\omega)
\\
\expvalequil{\psiLf {\omega' }^{A}
\psiRf {\omega }^{B}}  &=&
(1+ \me^{\beta \hbar\omega })^{-1} 
\sum_{B'=1}^{N}\delta^{AB'}R(\omega)^{B}_{B'}\,
2\pi \hbar\delta(\omega' +\omega)
.
\qquad
\label{eq:twopointequilLR}
\ena
The chiral energy-momentum currents are
\eqa
\TR(x-vt)_{A} &=&
\frac12 v T_{t}^{t}(x,t)_{A}  + \frac12 T_{t}^{x}(x,t)_{A} 
=
\frac{i}2 {:}\psiR(x,t)^{A} \partial_{t}\psiR(x,t)^{A}{:}
\qquad
\\
\TL (x+vt)_{A}  &=&
\frac12 v T_{t}^{t}(x,t)_{A} - \frac12 T_{t}^{x}(x,t)_{A}
=
\frac{i}2 {:}\psiL(x,t)^{A} \partial_{t}\psiL(x,t)^{A}{:}
\qquad
\ena
and their Fourier transforms are
\eqa
\TR(x-vt)_{A}
&=& \frac1{2\pi} \int_{-\infty}^{\infty} \dif \omega \,\, 
\me^{-i\omega (t-x/v)} \, \TRf{\omega}_{A}
\\[1ex]
\TL (x+vt)_{A}
&=&  \frac1{2\pi} \int_{-\infty}^{\infty} \dif \omega  \,\, 
\me^{-i\omega (t+x/v)} \, \TLf{\omega}_{A}
\ena
\eqa
\TRf{\omega}_{A}  &=&
\frac1{2\pi} \int \int \dif \omega _1  \dif \omega _2 
\,\delta(\omega _1+\omega_2-\omega)\,
\frac{\omega_2}2  {:}\psiRf {\omega _1}^{A}
\psiRf {\omega _2}^{A}{:}
\label{eq:TRmodes}
\\ 
\TLf{\omega}_{A}  &=&
\frac1{2\pi} \int \int  \dif \omega _1 \dif \omega _2
\,\delta(\omega _1+\omega_2-\omega) \,
\frac{\omega_2}2  {:}\psiLf {\omega _1}^{A}
\psiLf {\omega _2}^{A}{:}
\label{eq:TLmodes}
\:.
\ena
Their commutation relations with the fermion fields are
\eqa
\comm{\TRf{\omega' }_{A}}{ \psiRf {\omega }^{B}} &=&
-\delta_{A}^{B} \hbar (\frac12 \omega' + \omega )
\psiRf {\omega' +\omega }^{B} \\
\comm{\TLf{\omega' }_{A}}{ \psiLf {\omega }^{B}} &=& 
-\delta_{A}^{B}\hbar (\frac12 \omega'  + \omega ) 
\psiLf {\omega' +\omega }^{B}
\:.
\ena
The hamiltonian is
\eq
H_{0} = \sum_{A=1}^{N} \TRf{0}_{A} =\sum_{A=1}^{N} \TLf{0}_{A}
\:.
\en
Equation (\ref{eq:YentmodesB}) for 
$\Yent(\omega)_{AB}$ can now be evaluated,
using the entropy currents
\eqa
\JRf(\omega)_{A}&=&\kB \beta \TRf{\omega}_{A}  \\
\JLf(\omega)_{A}&=&\kB \beta \TLf{\omega}_{A}
\ena
and equations (\ref{eq:TRmodes}--\ref{eq:TLmodes}),
(\ref{eq:psiRpsiR}--\ref{eq:psiLpsiR}),
and~(\ref{eq:twopointequilRR}--\ref{eq:twopointequilLR}).
The result is
\eqa
\Yent(\omega)_{AB}
&=&
\frac{\kB^{2} \hbar \beta^{2}}{\omega}
\frac1{2\pi} \int \int  \dif \omega _1 \dif \omega _2
\,\delta(\omega _1+\omega_2-\omega) \,
\left [
R(\omega_{1})^{A}_{B} R(\omega_{2})^{A}_{B}
- \delta^{A}_{B}
\right ]
\nonumber \\
&& \qquad\qquad\qquad
\frac18 (\omega_{1}-\omega_{2})^{2}
\left (
\frac1{1+\me^{-\beta\hbar \omega_{2}}}
-
\frac1{\me^{\beta\hbar \omega_{1}} +1}
\right )
\:.
\label{eq:IsingYent}
\ena
The entropic ``capacitance'' is
\eq
\begin{split}
C_{SB} 
&= \lim_{\omega\rightarrow 0}
\frac{1}{i\omega}
\sum_{A=1}^{N}
\Yent(\omega)_{AB}
\\
&=
\frac{k}{T}
\frac1{2\pi i} \int \dif \eta
\left [
R^{-1} R'(\eta)
\right ]^{B}_{B}
\;
\frac{(\beta\hbar\eta)^{2}\me^{\beta \hbar \eta}}
{2 \left (\me^{\beta \hbar \eta}+1 \right )^{2}}
\:.
\label{eq:capacitanceIsing}
\end{split}
\en
The total entropic ``capacitance'' is
\eq
C_{S} = 
\frac{\dif s \hfill}{\dif T \hfill}
=
\sum_{B=1}^{N} C_{SB}
=
\frac{k}{T}
\frac1{2\pi i} \int \dif \eta \;
\Tr \left [ R^{-1} R'(\eta)\right ]
\;
\frac{(\beta\hbar\eta)^{2}\me^{\beta \hbar \eta}}
{2 \left (\me^{\beta \hbar \eta}+1 \right )^{2}}
\:.
\en
Integrating with respect to $T$ gives the junction entropy, 
up to a constant.  When $R(\omega)$ is independent of 
temperature, the integral can be done explicitly, giving
\eqa
s(T)-s(0) = 
\frac\kB{2\pi i} \int \dif \eta \;
\Tr \left [ R^{-1} R'(\eta) \right ]\;
\frac12
\left [
\frac{\beta \hbar \eta}{\me^{\beta \hbar \eta}+1 }
+\ln \left (
1+\me^{-\beta \hbar \eta}
\right )
\right ]
.
\quad
\ena
The information capacity of the junction is then
\eq
s(\infty)-s(0)
=
\kB \ln 2 \;
\frac1{2\pi i} \int \dif \eta \;
\frac12 
\Tr \left [ R^{-1} R'(\eta) \right ]
.
\label{eq:infocapacityIsing}
\en
The information capacity of the junction
thus consists of a half-bit for each pole of 
$R(\omega)^{A}_{B}$ in the lower half-plane
(or for each zero in the upper half-plane).

An elementary Ising junction
is a junction without substructure.
It has
a fermionic degree of freedom,
$\chi(t)^{A}$,
at the end of each wire,
satisfying
\eqa
\partial_{t}\chi(t) &=&
- \lambda
(\psiR +\psiL)(0,t)  \\
\lambda^{T}
\chi(t)  &=&
(\psiR -\psiL)(0,t)
 \ena
where $\lambda$ is a real $N\times N$ matrix.
The matrix elements $\lambda_{B}^{A}$ are the junction coupling 
constants.
Eliminating $\chi(t)$ gives boundary conditions
on the fermion fields:
\eq
(\partial_{t} +\lambda^{T}\lambda)\psiR(0,t)
 = (\partial_{t} -\lambda^{T}\lambda)\psiL(0,t)
\en
or, equivalently,
\eq
R(\omega) = \frac{\omega-i\lambda^{T}\lambda}
{\omega+i\lambda^{T}\lambda}
.
\label{eq:elemreflection}
\en
The information capacity of the elementary Ising junction
is given
immediately by (\ref{eq:infocapacityIsing}):
\eq
s(\infty)-s(0)
=
\frac{N}2 \kB\ln 2
\en
Equation (\ref{eq:capacitanceIsing}) for
the entropic ``capacitance'' is evaluated
by analytic continuation into the upper half-plane,
picking up the residues at the thermal poles.
The result is
\eq
C_{SB} =
\frac{\dif s \hfill}{\dif T \hfill}
= \kB^{2}\beta
F_{1}( \sigma  )_{B}^{B}
\en
where 
\eqa
\sigma &=& \frac{\beta \hbar \lambda^{T}\lambda}{2\pi} 
\\
F_{1}(\sigma) &=&
\sigma
-\sigma^{2}
\psi'( \frac12 + \sigma)
\\
\psi'(z) &=& (\ln \Gamma)''(z) =
\sum_{n=0}^{\infty} \frac1{(n+z)^{2}}
.
\ena
The total entropic ``capacitance'' is
\eq
C_{S} =
\frac{\dif s \hfill}{\dif T \hfill}
=\sum_{B=1}^{N}C_{SB} = \kB^{2}\beta
\;\Tr \;
F_{1} ( \sigma )
.
\en
Integrating with respect to $T$ gives
the junction entropy, up to a constant:
\eq
s(T) - s(0) =
\kB \Tr
\left [
F_{2}(\sigma)
-\sigma F_{2}'(\sigma)
\right ]
=
\kB 
\left (
1- \beta \frac{\partial\hfill}{\partial \beta}
\right )
\Tr
F_{2}(\sigma)
\en
where
\eq
F_{2}(\sigma)
= 
\sigma \ln \sigma  -\sigma - \ln \Gamma(\frac12 +\sigma) 
+\ln \sqrt{2\pi} 
.
\en
Therefore the boundary free energy, up to a constant, is 
\eq
\ln z(T) - \ln z(0) =
\Tr F_{2}
( \sigma )
.
\en
This agrees, in the case $N=1$,
with the direct calculation
of the boundary partition function
and the boundary entropy
of a bulk-critical Ising wire in a boundary
magnetic field.\cite{Chatterjee:1995sv,Konechny:2004dc}
The junction coupling constant, $\lambda$,
is the boundary magnetic field.

To have an explicit example of
the entropic admittance,
for illustration,
calculate $\Yent(\omega)_{AB}$ for the case $N=1$.
The wire labels, $A,B=1$, are suppressed.
Equation (\ref{eq:elemreflection}) for the reflection matrix is
\eq
R(\omega) = \frac{\omega  - i \lambda^{2}}{\omega  + i \lambda^{2}}
\en
In (\ref{eq:IsingYent}),
use the identity
\eq
R(\omega_{1})R(\omega_{2}) - 1 = 
\left (
\frac{\omega_{1}+\omega_{2}}{\omega_{2}-\omega_{1}}
\right )
\left [
R(\omega_{1})-R(\omega_{2})
\right ]
\en
then evaluate the integral by deforming the contour,
to obtain
the entropic admittance of
the boundary of a bulk-critical quantum Ising wire:
\eqa
\Yent(\omega )
&=&
\kB ^{2} \frac{\hbar}{2\pi}\beta^{2} \lambda^{2}i \omega
-\frac{1}{2}\kB ^{2}
\left (\frac{\hbar}{2\pi}\right )^{2}
\beta^{3} \lambda^{2}
\omega (\omega+2i\lambda^{2})
\nonumber \\
&&\qquad
\sum_{n=0}^{\infty}
\frac1{\left (n+\frac12+\frac{\beta \hbar \lambda^{2}}{2\pi}
\right )
\left ( n+\frac12+\frac{\beta \hbar \lambda^{2}}{2\pi}
-\frac{i \beta \hbar \omega}{2\pi}
\right )
}
\nonumber\\[1ex]
&=&
\kB ^{2} \frac{\hbar}{2\pi}\beta^{2} \lambda^{2}
i \omega
+
\kB ^{2} \frac{\hbar}{2\pi}\beta^{2} \lambda^{2}
\left ( \lambda^{2}-\frac{i\omega}2 \right )
\nonumber \\
&&\qquad
\left [
\psi\left(\frac12 +\frac{\beta\hbar}{2\pi}(\lambda^{2}-i\omega) 
\right )
-
\psi\left(\frac12 +\frac{\beta\hbar}{2\pi}\lambda^{2}
\right )
\right ]
.
\label{eq:IsingYent1wire}
\ena

\vskip2ex

\noindent
{\bf Acknowledgements}\\
I thank A.~Konechny for many discussions.
I thank
the members of an informal Rutgers seminar ---
S.~Ashok, A.~Ayyer, D.~Belov, E. Dell'Aquila,
B.~Doyon, and R.~Essig
--- for listening to a preliminary version of this work,
and for their helpful comments and questions.
I thank S.~Lukyanov for pointing towards some of the
condensed matter literature, leading in particular to
Ref.~\onlinecite{TsvelickWiegmann:1983}.
This work was supported by the Rutgers New High Energy 
Theory Center.

\newpage
\appendix
\addtocontents{toc}{\protect\setcounter{tocdepth}{0}}
\renewcommand{\thesection}{Appendix \arabic{section}:}
\section{The energy-momentum tensor at a junction}
\label{detail:boundaryenergy}

Consider $N$ wires connected at a junction.
Label the wires $A=1,\ldots , N$.
Each wire is parametrized by
a spatial coordinate, $x\ge 0$.
All the end-points, $x=0$, are identified to a single point, 
which is the junction.
The time coordinate is $t$;
the euclidean time is $\tau = i t$.
The euclidean space-time of each wire is
parametrized by the complex coordinate
$z = x+iv\tau = x-vt$,
$\bar z = x-iv\tau = x+vt$.
The space-time metric on wire $A$, in the bulk, is
\eq
g_{\mu\nu}(x)_{A} \dif x^{\mu}\dif x^{\nu}
= - v^{2}(\dif t)^{2} + (\dif x)^{2}
= \abs{\dif z}^{2} .
\en
At the junction, there is only the metric on time,
$g_{tt} = - v^{2}$.
Varying the space-time metric
gives the correlation functions of
the energy-momentum tensor:
\eq
\delta \expval{\cdots} = 
- \frac12 
\expval{ \frac{i}{\hbar} 
\int \dif t \,\,
\left [
\delta g_{tt}(t)T^{tt}(t)_\junct
+ \sum_{A=1}^{N} \int_{0}^{\infty} \dif x
\,\, 
\delta g_{\mu\nu}(x,t)_{A}
\,
T^{\mu\nu}(x,t)_{A}
\right ]
\cdots
}
.
\en
An infinitesimal localized reparametrization of space-time
is a collection of vector fields, $v^{\nu}(x,t)_{A}$,
one on each wire, all supported within a bounded region of 
space-time.
They must agree at the junction,
\eq
v^{\nu}(0,t)_{A} = v^{\nu}(t)_{\junct}
\:,
\en
and must leave the junction in place,
\eq
v^{x}(0,t)_{A} = v^{x}(t)_{\junct} = 0
.
\en
Such a reparametrization of space-time
is equivalent to changing the space-time metric by
\eqa
\delta g_{\mu\nu}(x,t)_{A} &=& \partial_{\mu}v_{\nu}(x,t)_{A}
+ \partial_{\nu}v_{\mu}(x,t)_{A} \\
\delta g_{tt}(t)_{\junct} &=& 2 \partial_{t}v_{t}(t)_{\junct}
.
\ena
The physics does not depend
on the parametrization of space-time,
so
\eq
\label{eq:conservation}
0=\int \dif t
\left [
\partial_{t}v^{t}(t)_{\junct} T^{t}_{t}(t)_{\junct}
+
\sum_{A=1}^{N}
\int_{0}^{\infty}\dif x \,\,
\partial_{\mu}v^{\nu}(x,t)_{A} T^{\mu}_{\nu}(x,t)_{A}
\right ]
\en
for all localized reparametrizations of space-time.
Integrating by parts gives the
local conservation of energy and momentum:
\eqa
0 &=& \partial_{\mu}T^{\mu}_{\nu}(x,t)_{A} \\
0 &=& \partial_{t}T^{t}_{t}(t)_{\junct}
+ \sum_{A=1}^{N} T^{x}_{t}(0,t)_{A}
.
\ena
Taking $v^{\mu}(x,t)$ to be infinitesimal time translation 
identifies the hamiltonian:
\eq
H_{0} = T^{t}_{t}(t)_{\junct} + \sum_{A=1}^{N}
\int_{0}^{\infty} \dif x \,\,  T_{t}^{t}(x,t)_{A}
.
\en
$T^{t}_{t}(x,t)_{A}$ is the bulk energy density in wire $A$.
$T^{t}_{t}(t)_{\junct}$ is the energy of the junction.

The departure from scale invariance is the trace of the 
energy-momentum tensor, $\Theta(x,t) = - T_{\mu}^{\mu}(x,t)$.
The junction contribution is conventionally written
$\theta(t)$.  That is, the junction energy is written
\eq
T_{t}^{t}(t)_{\junct} = - \theta(t)
\en
and conservation of energy at the junction is written
\eq
0 = - \partial_{t} \theta(t) +\sum_{A=1}^{N} T^{x}_{t}(0,t)_{A}
.
\en

\section{The chiral energy and entropy currents}
\label{app:chiralcommutators}

When the quantum wires are bulk-critical,
the bulk energy density and current
are linear combinations of chiral energy currents:
\eqa 
\TR(x,t)_{A} &=& \TR(x-vt)_{A} = \frac12 v T_{t}^{t}(x,t)_{A}
+ \frac12 T_{t}^{x}(x,t)_{A} \\
\TL(x,t)_{A} &=& \TL(x+vt)_{A} = \frac12 v T_{t}^{t}(x,t)_{A}
 - \frac12 T_{t}^{x}(x,t)_{A}
.
\ena
The bulk entropy current and density operators
are linear combinations of chiral entropy currents:
\eqa
\Jent(x,t)_{A} &=& \JR(x,t)_{A} - \JL(x,t)_{A} \\
\rhoent(x,t)_{A} &=& \frac1v\JR(x,t)_{A} + \frac1v\JL(x,t)_{A} 
\ena
where
\eqa
\JR(x,t)_{A}&=&\JR(x-vt)_{A} 
=\kB \beta \TR(x,t)_{A} -
\frac12 \kB \beta\expvalequil{ \TR(x,t)_{A} + \TL(x,t)_{A}}
\label{eq:JRB}\\
\JL(x,t)_{A} &=& \JL(x+vt)_{A}
= \kB  \beta \TL(x,t)_{A} -
\frac12 \kB \beta\expvalequil{ \TR(x,t)_{A} + \TL(x,t)_{A}}
\label{eq:JLB}
.
\ena
$\JR(x,t)_{A}$ flows to the right
$\JL(x,t)_{A}$ flows to the left,
both at the speed of ``light'', $v$:
$\JR(x,t+\delta t)_{A} = \JR(x-v\delta t,t)_{A}$,
$\JL(x,t+\delta t)_{A} = \JL(x+v\delta t,t)_{A}$.
The equilibrium expectation values in 
(\ref{eq:JRB}) and~(\ref{eq:JLB})
are subtracted so that $\expvalequil{\rhoent(x,t)_{A}} =0$.
Then
\eq
\expvalequil{\JR(x,t)_{A}} =
- \expvalequil{\JL(x,t)_{A}}
=
\frac12 \expvalequil{\Jent(x,t)_{A}} 
.
\en

The goal is to calculate the commutators
$\comm{\JR(x',t')_{A}}{\JR(x,t)_{B}}$
and $\comm{\JL(x',t')_{A}}{\JL(x,t)_{B}}$.
Because of the chirality of the currents, these commutators
of two currents of the same chirality
are completely determined by the equal-time commutators.
The equal-time commutators vanish if $A\ne B$,
so the commutators need only be calculated for $A=B$.
The labels $A,B$ are dropped during the calculation,
to simplify the notation, 
then restored at the end.

In 1+1 dimensional conformal field theory,
the chiral components of the energy-momentum tensor
are usually written $T(z)$, $\bar T(\bar z)$, where
\eqa
T(z) &=& 
-\frac{2\pi}{\hbar v^{2}} \TR(x-vt)
\label{eq:TTR}
\\
\bar T(\bar z) &=& 
-\frac{2\pi}{\hbar v^{2}} \TL(x+vt)
.
\label{eq:TbarTL}
\ena
The constant of proportionality, $-2\pi/\hbar v^{2}$,
will be confirmed shortly.
In euclidean space-time,
$T(z)$ and $\bar T(\bar z)$ 
satisfy operator product expansions
\eqa
T(z')\,T(z) &\sim& (z'-z)^{-4} \frac{c}{2}
+ (z'-z)^{-2} 2 T(z) 
+ (z'-z)^{-1}\partial T(z)  \\
\bar T(\bar z')\,\bar T(\bar z) &\sim& 
(\bar z'-\bar z)^{-4} \frac{c}{2}
+ (\bar z'-\bar z)^{-2} 2 \bar T(\bar z)
+ (\bar z'-\bar z)^{-1} \bar \partial \bar T(\bar z)
.
\ena
Write
the equal-time commutators as $\tau$-ordered 
operator products,
\eqa
\comm{T(x')}{T(x)} &=&
\tau\{T(x'+i\epsilon)\,T(x)\}
-\tau\{T(x'-i\epsilon)\,T(x)\}  \\
\comm{\bar T(x')}{\bar T(x)} &=&
\tau\{\bar T(x'-i\epsilon)\,\bar T(x)\}
-\tau\{\bar T(x'+i\epsilon)\,\bar T(x)\}
\:,
\ena
and evaluate them using the operator product expansions,
getting
\eqa
\frac{1}{2\pi i} \comm{T(x' )}{T(x )}
&=&
\frac{c}{12} \delta^{(3)}(x' -x )
+ (\partial_{x'}-\partial_{x})\,\left [ \delta(x' -x )\, T(x )
\right ]
\label{eq:TTcommeq}
\\[1ex]
\frac{-1}{2\pi i} \comm{\bar T(x' )}{ \bar T(x )}
&=&
\frac{c}{12} \delta^{(3)}(x' -x )
+ (\partial_{x'}-\partial_{x})\,\left [ \delta(x' -x )\, \bar T(x )
\right ]
.
\label{eq:TbarTbarcommeq}
\ena
Integrate over $x'$ to identify
the energy density as
\eq
T_{t}^{t}(x,t) =
-\frac{\hbar v}{2 \pi}
T(x-vt) - \frac{\hbar v}{2 \pi} \bar T(x+vt)
\:,
\en
confirming the constant of proportionality
in (\ref{eq:TTR}--\ref{eq:TbarTL}),
between $T(z)$ and $\TR(x-vt)$
and between $\bar T(\bar z)$ and $\TL(x+vt)$.
These equal-time commutation relations could 
have been derived 
equally well by combining conformal invariance
with the general equal-time commutation relations
derived in Appendix A of \onlinecite{EntFlowIJStatPhys}.
The equal-time commutation relations,
(\ref{eq:TTcommeq}) and~(\ref{eq:TbarTbarcommeq}),
can be written
\eqa
\frac{i}\hbar\comm{\TR(0,t')}{\TR(0,t)}
&=&
- \frac\hbar{2\pi} \frac{c}{12}
\partial_{t}^{3}\delta(t-t')
+(\partial_{t}-\partial_{t'})
\left [ \delta(t-t') \TR(0,t)
\right ]
\label{eq:TRTRcomm}
\\
\frac{i}\hbar\comm{\TL(0,t')}{\TL(0,t)}
&=&
- \frac\hbar{2\pi} \frac{c}{12}
\partial_{t}^{3}\delta(t-t')
+(\partial_{t}-\partial_{t'})
\left [ \delta(t-t') \TL(0,t)
\right ]
\label{eq:TLTLcomm}
.
\ena
The equilibrium expectation values of the commutators are
\eqa
\expvalequil{
\frac{i}\hbar\comm{\TR(0,t')}{\TR(0,t)}
}
&=&
- \frac\hbar{2\pi} \frac{c}{12}
\partial_{t}^{3}\delta(t-t')
+2\partial_{t}
\delta(t-t') \expvalequil{\TR(0,t)}
\label{eq:TRTRcommeq}
\\
\expvalequil{
\frac{i}\hbar\comm{\TL(0,t')}{\TL(0,t)}
}
&=&
- \frac\hbar{2\pi} \frac{c}{12}
\partial_{t}^{3}\delta(t-t')
+2\partial_{t}
\delta(t-t') \expvalequil{\TL(0,t)}
\label{eq:TLTLcommeq}
.
\ena

Now restore the wire labels, $A,B$.
Define the Fourier modes of the chiral energy-momentum 
currents by
\eqa
\TR(x,t)_{A}&=& \TR(x-vt)_{A}
 = \frac1{2\pi} \int_{-\infty}^{\infty} \dif \omega \,\, 
\me^{-i\omega (t-x/v)} \, \TRf{\omega}_{A}
\\[1ex]
\TL(x,t)_{A}&=& \TL(x+vt)_{A}
 =  \frac1{2\pi} \int_{-\infty}^{\infty} \dif \omega  \,\, 
\me^{-i\omega (t+x/v)} \, \TLf{\omega}_{A}
\ena
The commutation relations,
(\ref{eq:TRTRcomm}) and~(\ref{eq:TLTLcomm}),
are equivalent to
\eqa
\comm{\TRf{\omega'}_{B}}{\TRf{\omega}_{A}}
&=&
\delta_{AB} \left [
-\frac{c}{12} \hbar^{2}\omega^{3}\delta(\omega'+\omega)
+\hbar (\omega'-\omega) \TRf{\omega'+\omega}_{A}
\right ]
\\
\comm{\TLf{\omega'}_{B}}{\TLf{\omega}_{A}}
&=&
\delta_{AB} \left [
-\frac{c}{12} \hbar^{2}\omega^{3}\delta(\omega'+\omega)
+\hbar (\omega'-\omega) \TLf{\omega'+\omega}_{A}
\right ]
.
\ena
Conservation of energy at the junction, 
(\ref{eq:energy-inflow}), implies that
\eq
\sum_{A} \TRf{0}_{A} - \sum_{A} \TLf{0}_{A} = 0
.
\en
The hamiltonian acts by
\eqa
\comm{H_{0}}{ \TRf{\omega}_{A}} &=& -\hbar \omega  \TRf{\omega}_{A}
\\
\comm{H_{0}}{ \TLf{\omega}_{A}} &=& -\hbar \omega  \TLf{\omega}_{A}
\ena
so the hamiltonian is
\eq
H_{0} = \sum_{A} \TRf{0}_{A} = \sum_{A} \TLf{0}_{A}
.
\en
A straightforward calculation shows these expressions for 
the hamiltonian to be equal to
the spatial integral of the energy density.

The equilibrium expectation values of the commutators are
\eqa
\expvalequil{
\comm{\TRf{\omega'}_{B}}{\TRf{\omega}_{A}}
}
&=&
\delta_{AB} \left [
-\frac{c}{12} \frac{\hbar}{2\pi} \omega^{3}
- 2 \omega \expvalequil{\TR(0,t)_{A}}
\right ]
2\pi \hbar \delta(\omega'+\omega)
\\
\expvalequil{
\comm{\TLf{\omega'}_{B}}{\TLf{\omega}_{A}}
}
&=&
\delta_{AB} \left [
-\frac{c}{12} \frac{\hbar}{2\pi} \omega^{3}
- 2 \omega \expvalequil{\TL(0,t)_{A}}
\right ]
2\pi \hbar \delta(\omega'+\omega)
.
\ena
The equilibrium one-point functions,
\eqa
\expvalequil{\TR(x,t)_{A}} &=& \expvalequil{\TR(x-vt)_{A}}
\\
\expvalequil{\TL(x,t)_{A}} &=& \expvalequil{\TL(x+vt)_{A}}
\:,
\ena
are independent of $t$, so are constant in $x$.
They can be evaluated far from the junction,
where the system is conformally invariant.
When no equilibrium currents are flowing,
when $\expvalequil{T_{t}^{x}(0,t)_{A}}=0$,
the equilibrium bulk energy density
is\cite{BLOTECARDYNIGHTINGALE,AFFLECK1986}
\eq
\expvalequil{T_{t}^{t}(x,t)_{A}} 
=  \frac{c}{12} \frac{2\pi}{\hbar v\beta^{2}} 
.
\en
Equivalently,
\eq
\expvalequil{\TR(x,t)_{A}} = \expvalequil{\TL(x,t)_{A}}
= \frac{c}{24} \frac{2\pi}{\hbar \beta^{2}}
\label{eq:expvalTRTL} 
.
\en
The equilibrium expectation values of the commutators
are then
\eqa
\expvalequil{\comm{\TRf{\omega'}_{B}}{\TRf{\omega}_{A}}}
&=&
- \delta_{AB} \delta(\omega'+\omega)
\frac{c}{12} \hbar^{2}
\left [ \omega^{3} +
\left ( \frac{2\pi}{\hbar \beta}
\right )^{2} \omega
\right ] 
\\
\expvalequil{\comm{\TLf{\omega'}_{B}}{\TLf{\omega}_{A}}}
&=&
- \delta_{AB} \delta(\omega'+\omega)
\frac{c}{12} \hbar^{2}
\left [ \omega^{3} +
\left ( \frac{2\pi}{\hbar \beta}
\right )^{2} \omega
\right ] 
\label{eq:TTnocurrent}
.
\ena
Bulk conformal invariance requires that this commutator 
vanish at $\omega=2\pi i/\hbar \beta$.
This condition gives another way to derive
the equilibrium energy density.

\newpage
\bibliographystyle{spmpsci}       
\bibliography{entropy}

\end{document}